# The Benefit of Being Flexible in Distributed Computation

Linqi Song, Sundara Rajan Srinivasavaradhan, Christina Fragouli
University of California, Los Angeles CA 90095, USA
Email: {songlinqi,sundar,christina.fragouli}@ucla.edu

*Abstract*—In wireless distributed computing, networked nodes perform intermediate computations over data placed in their memory and exchange these intermediate values to calculate function values. In this paper we consider an asymmetric setting where each node has access to a random subset of the data, i.e., we cannot control the data placement. The paper makes a simple point: we can realize significant benefits if we are allowed to be "flexible", and decide *which node computes which function*, in our system. We make this argument in the case where each function depends on only two of the data messages, as is the case in similarity searches. We establish a percolation in the behaviour of the system, where, depending on the amount of observed data, by being flexible, we may need no communication at all.

## I. INTRODUCTION

Wireless distributed computing divides a computation over a set of data into tasks that are solved by a cluster of networked nodes. One way to view such systems is as trading computation with communication cost. A single server may collect all data and perform all computations; but it may also outsource most of the computation and spend communication resources to distribute data and collect the computation intermediate values. Such exchanges may enable to operate with lower latency or lower battery consumption; the decision on how to operate is determined by examining the computation and the communication cost in each case. Accordingly, recent work in the literature has started exploring communication vs. computation trade-off [1], where the main techniques stem from network coding and index coding [2], [3].

Data locality or data availability is an important consideration in distributed computing, where we would like the subtasks to be assigned to distributed computing nodes that contain as much of the data needed to compute the subtasks as possible [4]. The goodness of data locality is defined as the percentage of nodes that have all the data to compute their subtasks. Previous work looked at how to assign subtasks that minimize the uncoded communication cost between nodes [4], [5]. However, when coding techniques are used, a naturally interesting question is to how we could assign the subtasks for minimal communication.

In this paper, we focus on the case of random data placement, where each node has collected a random subset of the data that it can use for processing - we cannot control the original data placement. This is plausible in wireless, where the processing nodes are often mobile, equipped with sensors, and can collect (potentially overlapping) data: we may want to process these data in a time-critical fashion and upload information to a remote server for portable scientific computing; we may want to combine different views of the data (say videos) for virtual reality applications; we may want time-sensitive processing for distributed tasks, such as distributed online recommendations [6]; or we may have a crowdsourcing application where we want to locally process for privacy considerations - nodes may be willing to share function values, but not the raw data [7].

We operate under the framework of MapReduce [4]. The first stage "Map" is intra-node data processing and local computation, followed by the data shuffling which collects intermediate computation results with the same "key" to the same computing node, while the second stage "Reduce" computes each function based on these local computation results with the same "key". As observed in [8], [9] and [10], data shuffling is a limiting factor in the runtime performance of many distributed computing tasks. Moreover, we may want to calculate functions that depend on only a subset of the data, that satisfy a certain attribute (timestamps, geographical area, origin etc).

The asymmetry (in terms of the function-data relationship) in our framework makes clear the following point: we can gain significant benefits, in terms of communication, by being "flexible", and leveraging on the fact that in distributed computing, often we not only know in advance what the functions we want to compute are, but also decide *which of the nodes in our system computes each function*. For instance, we may eventually upload the computation outcome to a centralized server, and thus what is important is to efficiently compute all functions - not which node does what. In the symmetric scenario where each function depends on all data, it makes no difference which node computes which function; in contrast, in asymmetric scenarios, we find a significant difference: depending on the amount of data nodes randomly observe, nodes may not need to communicate at all; in contrast, when functions are allocated independently of the random data realization, this is almost never the case. We establish a percolation in the behaviour of the system, where, depending on the amount of observed data, by being flexible, we may need no communication when the probability to observe a data message is above some threshold or we may need significant

This work was supported in part by the National Science Foundation under grant number 1527550.

communication when this probability is below the threshold, even when allowed to use coded broadcast transmissions.

Each function could, in principle, depend on an arbitrary number of data messages. We start from the simplest case where each function depends on only two of the data messages, which is still useful in practice, such as in similarity searches. As a concrete example, consider the task of finding "common friends", a task that is regularly computed in social networks, like Facebook, where friendships dynamically change. Consider that we have a social network, where the friendships between users are stored as messages: the message $b_A$ is user $A$'s set of friends $(A, \{B, C, D\})$, the message $b_B$ is user $B$'s set of friends $(B, \{A, D, E\})$, similarly, $b_C = (C, \{A, E\}), b_D = (D, \{A, B, F\}), b_E = (E, \{B, C, F\}), b_F = (F, \{D, E\})$. These 6 messages can be distributed among computing nodes, for example four nodes as follows: node 1 has $\{b_A, b_C, b_E\}$, 2 has $\{b_B, b_D, b_F\}$, 3 has $\{b_B, b_E, b_F\}$, and 4 has $\{b_A, b_C, b_D\}$. Assume we want to find the common friends of the following $K = 3$ pairs: $\{\{A, B\}, \{B, C\}, \{D, E\}\}$. The data shuffling scheme needs to gather the messages needed to compute these $K$ "common friends" to some $K$ nodes. For example, node 4 can broadcast $b_A$ and $b_C + b_D$; then node 1 can decode message $b_D$ using the transmission $b_C + b_D$ and its local data $b_C$ and then compute the common friends of $\{D, E\}$; similarly, node 2 can decode message $b_C$ and compute the common friends of $\{B, C\}$, and node 3 can decode message $b_A$ and compute the common friends of $\{A, B\}$. Such a "common friends" type tasks pervade a number of computing tasks today; for example, in recommender systems [6], to evaluate the similarity between two users, we need to find those liked videos that are common to the two, or the products bought by both.

The paper is organized as follows. Section II introduces our notation and problem formulation; Section III considers the case where we can flexibly decide which node calculates which function, while Section IV looks at the case where we fix the allocation of functions to nodes.

## II. PROBLEM FORMULATION

Throughout the paper, we will use $[y]$ (with $y$ a positive integer) to denote the set $\{1, 2, \ldots, y\}$ and $|Y|$ to denote the cardinality of set $Y$.

We consider a distributed computing system with $n$ computing nodes $c_1, c_2, \ldots, c_n$, and $m$ data messages, $b_1, b_2, \ldots, b_m$ in some finite field $\mathbb{F}_q$. We will interchangeably use $b_j$ and message $j \in [m]$ to refer to messages, and similarly $c_i$ or node $i$ to refer to nodes. We assume that each node $c_i$, $i \in [n]$, has been allocated message $b_j$, $j \in [m]$ with allocation probability $p$ independently and uniformly across nodes and messages. The allocation probability $p$ is interpreted as the probability that a computing node contains a message. This is motivated by the fact that sometimes, the distributed nodes collect data separately and randomly, such as wireless sensors collect data within some region. Therefore, the relationship between the $m$ messages and the $n$ nodes can be represented by a random bipartite graph $B(m, n, p)$[1]. The messages allocated to node $c_i$ are referred to as side information, and indexed by a set $S_i \subseteq [m]$.

We assume we have $K \leq n/2$ computation functions to compute and each function takes two data messages as input, namely, $f_1(b_{11}, b_{12}), f_2(b_{21}, b_{22}), \ldots, f_K(b_{K1}, b_{K2})$, for two different messages $b_{k1}, b_{k2} \in \{b_j | j \in [m]\}$, for all $k \in [K]$. The functions and their dependency on messages is known in advance. Since we can discard a message $b_j$ if no function relies on it, we always assume that each message is useful to compute some function(s), but no more than $\Theta(1)$ functions. In distributed computation, each function $f_k(b_{k1}, b_{k2})$ can be calculated as

$$f_k(b_{k1}, b_{k2}) = h_k(v_{k1}^k, v_{k2}^k) = h_k(l_k^1(b_{k1}), l_k^2(b_{k2}))$$

where $h_k$ is a function of $v_{k1}^k, v_{k2}^k$ and functions $v_{k1}^k = l_k^1(b_{k1})$, $v_{k2}^k = l_k^2(b_{k2})$ are intermediate results of the distributed computation[2]. Given the messages in $S_i$, node $i$ can calculate the values $v$ for all functions that take attributes in $S_i$. We use $\mathcal{V}_i$ to denote the set of these values.

The data shuffling phase ensures nodes collect the intermediate values they need by communicating with each other. We assume that a set of nodes make $T$ broadcast transmissions, $x_1, x_2, \ldots, x_T$, where a broadcast transmission made by node $i \in [n]$ is a linear combination of the messages in $\mathcal{V}_i$. We will refer to transmissions as *uncoded*, if they contain only one message (do not use coding).

We say that $f_k$ *is covered by a node* $i$ if $i$ has both messages $b_{k1}$ and $b_{k2}$ and thus can calculate the values $v_{k1}^k$ and $v_{k2}^k$; we say that $f_k$ is covered by a node $i$ and a set of $h$ transmissions $x_{t1}, x_{t2}, \ldots, x_{th}$ if $i$ can compute $f_k$ given its own messages and the partial results it decodes from the transmissions.

We aim to minimize the number of broadcast transmissions $T$, such that there exist a set of $K$ different computing nodes $i_1, i_2, \ldots, i_K$ that can compute functions $f_1, f_2, \ldots, f_K$, respectively. That is, we decide which function is computed by which of the nodes so as to minimize the communication overhead. In the random bipartite graph formulation, we ask what is the typical behaviour of the minimum number of transmissions $T$ as the parameters $n, m, K$ tend to infinity.

## III. FLEXIBLE FUNCTION ASSIGNMENT

Assume we can select which node calculates which function. We start by defining a threshold property.

**Definition 1.** *The function* $p_{th}(m, n, K)$ *is a threshold for the random graph instance* $B(m, n, p)$ *regarding an increasing property* $\mathcal{P}$, *if*

$$\lim_{m,n,K \to \infty} \Pr\{B(m, n, p) \text{ satisfies } \mathcal{P}\}$$
$$= \begin{cases} 0, & p/p_{th}(m, n, K) \to 0, \\ 1, & p/p_{th}(m, n, K) \to \infty. \end{cases}$$

---
[1]This is essentially a random $m \times n$ binary matrix where each element is 1 independently with probability $p$ and 0 otherwise.
[2]The data shuffling phase may transmit these intermediate values $v_{k1}^k, v_{k2}^k$.

We then prove a "percolation" or "threshold" property: when the probability $p$ is above a threshold, we do not need shuffling; i.e., we can find a "matching" between nodes and functions, such that each node almost surely has the data needed to compute its allocated function, with no communication. Theorem 1 follows from Lemmas 1 and 2.

**Theorem 1.** *Given $K \leq n/2$, $K \leq \binom{m}{2}$, and $\sqrt{n} = K^{1-\epsilon} o(1)$ for some fixed $\epsilon > 0$, the function $p_{th}(m,n,K) = \sqrt{\frac{\log(K)}{n}}$ is a threshold for the random graph instance $B(m,n,p)$ regarding the property that no data shuffling is needed.*

### A. Allocation Probability Above Threshold

We here look at the case where no communication is needed, provided we decide which node calculates which function.

**Lemma 1.** *If the allocation probability $p$ satisfies $p = \sqrt{\frac{\log(K)}{n}} \omega(1)$, then data shuffling is not needed almost surely, i.e., with probability at least $1 - (\frac{1}{K})^{\frac{1}{2}\omega(1)-1} = 1 - o(1)$, we can find in polynomial time an allocation of functions to nodes, such that each function is covered.*

*Proof.* Since each message is useful for no more than $\Theta(1)$ functions, we can separate the functions into $\Theta(1)$ groups such that any two functions in the same group depend on different messages. Here we only consider 1 group case and this can be applied to $\Theta(1)$ groups. First, we can see that the probability that a node $i$ has both messages $b_{k1}$ and $b_{k2}$ to compute the function $f_k$ is $p^2$. Then the probability that the function $f_1$ is not covered by any of the $n$ nodes (i.e., $b_{11}$ and $b_{12}$ are not cached simultaneously at any node) is:

$$\Pr\{f_1 \text{ is not covered by } n \text{ nodes}\} = (1-p^2)^n.$$

Assume we assign $f_1$ to one node that covers it. The probability that $f_2$ is not covered by $n-1$ nodes (i.e., $b_{21}$ and $b_{22}$ are not cached simultaneously at a given $n-1$ nodes) is:

$$\Pr\{f_2 \text{ is not covered by } n-1 \text{ nodes}\} = (1-p^2)^{n-1}.$$

Similarly, the probability that $f_k$ (for $k = 3, \ldots, K$) is not covered by $n-k+1$ nodes (i.e., $b_{k1}$ and $b_{k2}$ are not cached simultaneously at a given $n-k+1$ nodes) is:

$$\Pr\{f_k \text{ is not covered by } n-k+1 \text{ nodes}\} = (1-p^2)^{n-k+1}.$$

Hence, the probability that some function among all $K$ functions cannot be computed directly is

$$\Pr\{\exists k \in [K], f_k \text{ cannot be computed directly}\}$$
$$\leq \Pr\{f_1 \text{ is not covered by } n \text{ nodes}\} + \cdots +$$
$$\Pr\{f_K \text{ is not covered by } n-K+1 \text{ nodes}\}$$
$$\leq (1-p^2)^n (1 + \frac{1}{1-p^2} + \ldots + \frac{1}{(1-p^2)^{K-1}})$$
$$\leq K(1-p^2)^{n-K} \leq K e^{-\omega(1)(1-K/n)\log(K)}$$
$$\leq (\frac{1}{K})^{\frac{n-K}{n}\omega(1)-1} \leq (\frac{1}{K})^{\frac{1}{2}\omega(1)-1} = o(1).$$

To allocate functions to nodes, we can create a bipartite graph that lists all $n$ computing nodes on one side and all $K$ functions on the other side, and connect a function with a computing node if and only if the node has both of the messages required to compute this function. We can then in polynomial time solve the maximum bipartite graph matching problem (see [11] and [12], for example). □

### B. Allocation Probability Below Threshold

We here calculate the number of uncovered functions in the case where $p < p_{th}(m,n,K)$ and provide a lower bound on the amount of communication needed.

**Lemma 2.** *Given $\sqrt{n} = o(K^{1-\epsilon})$ for some fixed $\epsilon > 0$, if the allocation probability satisfies $p = \sqrt{\frac{\log(K)}{n}} o(1)$, then the number of uncovered functions is $K^{1-o(1)}/2$ almost surely, i.e., with probability at least $1 - \exp(-\frac{K^{2-o(1)}}{8n}) = 1 - o(1)$.*

*Moreover, if the allocation probability $p \leq \sqrt{\frac{1}{n}}$, then the number of uncovered functions is $\frac{K}{2e}$ almost surely, i.e., with probability at least $1 - \exp(-\frac{K^2(1-o(1))}{8e^2 n}) = 1 - o(1)$.*

*Proof.* We define $Y^C$ to be the maximum number of functions that are covered by different nodes, i.e., both $b_{k1}$ and $b_{k2}$ for the functions $f_k$ are in the side information set of some distinct nodes. We also define $Y = K - Y^C$ as the minimum number of uncovered functions. From the proof of Lemma 1, we know that the probability that a function $f_k$ is not covered is:

$$\Pr\{f_k \text{ is not covered by } n \text{ nodes}\} = (1-p^2)^n. \quad (1)$$

Therefore, the average number of functions that are not covered is

$$\begin{aligned}\mathbb{E}Y &\geq K(1-p^2)^n \geq K(1 - \frac{o(1)\log(K)}{n})^n \\ &\stackrel{(a)}{\geq} K e^{-o(1)\log(K)}(1 - \frac{o(1)\log^2(K)}{n}) = K^{1-o(1)},\end{aligned} \quad (2)$$

where $(a)$ follows from $(1+x/n)^n \geq e^x(1-x^2/n)$. Next, we show that the number $Y$ is distributed tightly around $\mathbb{E}Y$. To show this, we first use a "vertex exposure" process to form a martingale based on the random graph $B(m,n,p)$ [13], [14]. Specifically, we label the $n$ computing nodes in the random bipartite graph as $1, 2, \ldots, n$ and denote by $Z_l$ the random variable to indicate whether the vertex $l$ is exposed in the random graph, i.e., $Z_l = 1$ if the $l$-th node is present in the graph and $Z_l = 0$ otherwise. Define $Y_l = \mathbb{E}[Y|Z_1, Z_2, \ldots, Z_l]$ as a sequence of random variables for $l = 1, 2, \ldots, n$, then $\{Y_l\}$ is a Doob martingale and $Y_n = Y$. Observe that when we add (or remove) a node, the minimum number of uncovered functions can decrease (or increase) at most 1. In fact, after adding some messages to node $i$'s side information, if $Y$ decreases by 2, then without counting on node $i$, the number $Y$ will decrease by at least 1. This contradicts the fact that $Y$ is the minimum number of uncovered functions. We conclude that the function $Y$ is 1-Lipschitz: if two random graph realization $B_0$ and $B_1$ differ at one vertex $l$, namely, the subgraph induced by vertices of $B_0$ other than $l$ is the same as that induced by vertices of $B_1$ other than $l$, and the function $Y$ satisfies $|Y(B_0) - Y(B_1)| \leq 1$. Therefore, we can use the

Azuma's inequality

$$\Pr\{\mathbb{E}[Y] - Y \geq a\} \leq \exp(-\frac{a^2}{2n}), \text{ for } a > 0 \quad (3)$$

to get

$$\Pr\{Y \leq 0.5K^{1-o(1)}\} \leq \Pr\{\mathbb{E}[Y] - Y \geq 0.5K^{1-o(1)}\}$$
$$\leq \exp(-\frac{K^{2-o(1)}}{8n}).$$

The first part of the theorem is proved. For the second part of the theorem, we use similar techniques. The average number of functions that are not covered is

$$\mathbb{E}Y \geq K(1-p^2)^n \geq K(1-\frac{1}{n})^n \geq Ke^{-1}(1-\frac{1}{n})). \quad (4)$$

Thus, we can also bound the number $Y$ using the martingale construction and Azuma's inequality:

$$\Pr\{Y \leq \frac{K}{2e}\} \leq \Pr\{\mathbb{E}[Y] - Y \geq \frac{K}{2e}(1-o(1))\}$$
$$\leq \exp(-\frac{K^2(1-o(1))}{8e^2 n}). \quad (5)$$

$\square$

Note that for the above lemma, the claim holds for $\sqrt{n} = K^{1-\epsilon}o(1)$, for example, $K = \Theta(n)$.

Next, we show that for $p < p_{th}(m,n,K)$, we need a non-trivial amount of communication in the data shuffling phase, even if we are allowed to use coding.

**Theorem 2.** *For allocation probability $p \leq p_{th}(m,n,K)$, the optimal number of (coded) broadcast transmissions is almost surely lower bounded by $\Omega(T_{un})$, where $T_{un}$ is the minimum number of uncoded broadcast transmissions we would need to make, if after the random data placement realization, we broadcast missing messages $b_j$, $j \in [m]$ so that any $K$ out of the $n$ nodes can compute the $K$ functions.*

The proof outline of this theorem consists of two major steps. First, similar to [15], we need to categorize all $\Omega(T_{un}) \times m$ matrices into a smaller number of "clusters" such that matrices in each "cluster" have similar decoding ability, i.e., all matrices in each "cluster" can satisfy a problem instance with low probability. Second, we count all the number of matrix "clusters" and prove that in total the probability is low for all "clusters" to satisfy a problem instance. The complete proof of this theorem can be found in Appendix A. Intuitively, the reason why coding does not help, is because the side information is limited.

### C. Outage Probability

If the allocation probability is too low, it is probable that none of the cluster nodes observes some of the data, and thus the computation of functions that depend on this data becomes impossible. We show that this outage probability is $p_{out} = \frac{\log(m)}{n}$, where for $p \geq p_{out}(1+\epsilon)$, the probability that some message is missing is almost surely 0, while for $p \leq p_{out}(1-\epsilon)$, this probability is almost surely 1, for any fixed $\epsilon > 0$.

The probability that a message $b_j$ is not cached in one of the $n$ nodes is $(1-p)^n$. Then, the probability that a message $b_j$ is not missing is $1-(1-p)^n$. So the probability that some message is missing is $p_{miss} \triangleq 1 - [1-(1-p)^n]^m$. Hence, we have for $p \leq p_{out}(1-\epsilon)$,

$$\begin{aligned}p_{miss} &\geq 1 - [1 - e^{-np}(1-np^2)]^m \\ &\geq 1 - [1 - m^{-1+\epsilon}(1-o(1))]^m \\ &\geq 1 - e^{-m^\epsilon(1-o(1))} = 1 - o(1);\end{aligned} \quad (6)$$

and for $p \geq p_{out}(1+\epsilon)$,

$$\begin{aligned}p_{miss} &\leq 1 - [1 - m(1-p)^n] \leq me^{-np} \\ &\leq me^{-(1+\epsilon)\log(m)} = (\frac{1}{m})^\epsilon = o(1).\end{aligned} \quad (7)$$

## IV. BENEFIT WRT FIXED FUNCTION ASSIGNMENT

In this section, we calculate how much we have gained by being flexible, as compared to fixing the function assignment. We compare with the case where each function $f_k$ is assigned to some $n_k = \Theta(1)$ disjoint computing nodes, independently of the data placement. When $n_k = 1$ for all $k$, t we assign each function to one specific node. Let $C \triangleq \sum n_k/K$ to be the average number of computing nodes to compute each function.

**Theorem 3.** *Given a random graph instance $B(m, \Theta(K), p)$, the threshold probability satisfies $p_{th}(m, \Theta(K), K) = 1 - (\frac{\log(K)}{K})^{1/C} = 1 - o(1)$ with respect to the property that data shuffling is not needed for fixed function assignments.*

*Proof.* We now consider the case when $p \leq p_{th}$. It suffices to prove that data shuffling is almost always needed for this case. We observe that the probability that $f_k$ is covered by at least one of its corresponding $n_k$ worker nodes is

$$\Pr\{f_k \text{ is covered}\} = 1 - (1-p^2)^{n_k}.$$

Now, the probability that all functions can be computed is

$$\Pr\{\forall k, f_k \text{ can be computed directly}\}$$
$$= \prod_{k=1}^{K} \Pr\{f_k \text{ is covered}\} = \prod_{k=1}^{K} \left(1 - (1-p^2)^{n_k}\right)$$
$$\stackrel{(a)}{\leq} \left(1 - \sum_{k=1}^{K} \frac{(1-p^2)^{n_k}}{K}\right)^K \stackrel{(b)}{\leq} (1-(1-p^2)^C)^K$$
$$\leq (1-(1-p_{th})^C)^K \leq (1 - \frac{\log(K)}{K})^K \leq \frac{1}{K}$$

where $(a)$ follows from Arithmetic Mean – Geometric Mean (A.M-G.M) inequality applied to the product term, and $(b)$ follows, again, from A.M-G.M applied to the summation term inside the parenthesis. $\square$

### A. Uncoded transmissions

Let us consider that only one node is designated to each function. We would like to compute the number of uncoded transmissions needed in this case.

**Theorem 4.** *Given a random graph instance $B(m,n,p)$, the threshold function satisfies $p_{th}(m,n,K) = 1 - \frac{1}{K^\epsilon}$ (for any fixed $\epsilon > 0$) with respect to the property that $\Theta(K^{1-\epsilon})$ uncoded transmissions are needed for a successful computation of all functions.*

*Proof.* Observe that for a function $f_k$, we may need to make 2, 1, or 0 uncoded transmissions. The probabilities are $(1-p)^2$, $2p - p^2$, and $p^2$, since a node needs to have two messages to compute its function. Therefore the expected number of uncoded transmissions is $K(2p - p^2 + 2(1-p)^2) = K(2 - 2p + p^2)$. Using Chernoff's bound, we get that the probability for the number of transmissions $X$ taking a value less than or equal to $(1-\delta)K(2 - 2p + p^2)$ (for $\delta \in (0,1)$) is,

$$\Pr\{X \leq (1-\delta)K(2-2p+p^2)\}$$
$$\leq e^{-K(1-p+p^2/2)\delta^2/2} \leq e^{-\Theta(K^{1-\epsilon})}$$

□

### B. The Benefit of Being Flexible

We now summarize and compare our results for the flexible and fixed function assignments and show the benefits of being flexible in function assignment.

1) $p < p_{out}$: in this regime, the distributed computing task is not possible irrespective of function assignments.
2) $p_{out} < p = \sqrt{\frac{\log(K)}{n}} o(1)$: in this regime, we may need significant amount of communication, but there is no significant coding gain for flexible function assignments. The uncoded transmissions can be as large as $\Theta(K)$.
3) $\sqrt{\frac{\log(K)}{n}} \omega(1) = p < 1 - \frac{\log(K)}{K}$: this regime spans most of the allocation probability region, ranging from $o(1)$ to $1 - o(1)$, and underscores the benefit of the extra degree of freedom (in the form of flexibility in function assignment). A fixed function assignment scheme almost surely needs $\Theta(K)$ transmissions for constant allocation probability, while flexibility in function assignment almost surely requires no communication.

## V. RELATED WORK

In distributed computing, data shuffling or communication among computing nodes form a major bottleneck for the runtime performance. Recently, as distributed mobile computing is increasingly attracting interest, coding techniques (e.g., network coding [2], index coding [3]) can be used to reduce the communication cost in distributed computing [16], [17], [1], [18], [19]. Closer to ours is work that helps improve the communication efficiency in distributed systems. The work of [17], [18], [19] has studied the "master-workers" distributed computing model and show that coding can help to reduce communication cost in data shuffling, for example, in distributed machine learning applications. The work in [16], [1] has established that leveraging of broadcasting and coding can improve the communication efficiency of such systems. However, creating coding opportunities is enabled by allocating data in specific patterns across the nodes [1] (in fact, similar data allocations have also been used for distributed caching [20]). In contrast, in our work we assume random data allocation, and thus the trade-off curve in [1] does not apply. We also do not assume predetermined message requests, but we decide where to send intermediate function values, which makes the problem different from conventional index coding techniques; intermediate values are only useful to one node, and thus to create side information, we need to make computations. In conventional index coding problems, the messages to be transmitted to receivers and the side information of receivers are fixed and the problem is to design appropriate schemes to minimize the number of broadcast transmissions. However, when using coding schemes in distributed computing, such as in [16], [17], [19] and this work, we not only need to design transmission schemes, but also need to design the data placement and schedule the task assignment.

## REFERENCES

[1] S. Li, M. A. Maddah-Ali, and A. S. Avestimehr, "Fundamental tradeoff between computation and communication in distributed computing," in *IEEE International Symposium on Information Theory (ISIT)*, 2016.
[2] S.-Y. Li, R. W. Yeung, and N. Cai, "Linear network coding," *Information Theory, IEEE Transactions on*, vol. 49, no. 2, pp. 371–381, 2003.
[3] Z. Bar-Yossef, Y. Birk, T. Jayram, and T. Kol, "Index coding with side information," *IEEE Transactions on Information Theory*, vol. 57, no. 3, pp. 1479–1494, 2011.
[4] J. Dean and S. Ghemawat, "Mapreduce: simplified data processing on large clusters," *Communications of the ACM*, vol. 51, no. 1, pp. 107–113, 2008.
[5] Z. Guo, G. Fox, and M. Zhou, "Investigation of data locality in mapreduce," in *Cluster, Cloud and Grid Computing (CCGrid), 2012 12th IEEE/ACM International Symposium on*. IEEE, 2012, pp. 419–426.
[6] F. Ricci, L. Rokach, and B. Shapira, *Introduction to recommender systems handbook*. Springer, 2011.
[7] E. Toch, "Crowdsourcing privacy preferences in context-aware applications," *Personal and ubiquitous computing*, vol. 18, no. 1, pp. 129–141, 2014.
[8] M. Chowdhury, M. Zaharia, J. Ma, M. I. Jordan, and I. Stoica, "Managing data transfers in computer clusters with orchestra," in *ACM SIGCOMM Computer Communication Review*, vol. 41, no. 4, 2011, pp. 98–109.
[9] F. Ahmad, S. T. Chakradhar, A. Raghunathan, and T. N. Vijaykumar, "Tarazu: Optimizing mapreduce on heterogeneous clusters," *SIGARCH Comput. Archit. News*, vol. 40, no. 1, pp. 61–74, Mar. 2012.
[10] Y. Guo, J. Rao, and X. Zhou, "iShuffle: Improving hadoop performance with shuffle-on-write," in *10th International Conference on Autonomic Computing (ICAC 13)*, 2013.
[11] R. Diestel, *Graph theory; 2nd ed.* Heidelberg: Springer, 2000.
[12] M. Mucha and P. Sankowski, "Maximum matchings via gaussian elimination," in *Proceedings. 45th Annual IEEE Symposium on Foundations of Computer Science*, 2004, pp. 248–255.
[13] B. Bollobás, *Random graphs*. Cambridge Studies in Advanced Mathematics 73, 2001.
[14] N. Alon and J. H. Spencer, *The probabilistic method*. John Wiley & Sons, 2004.
[15] A. Golovnev, O. Regev, and O. Weinstein, "The minrank of random graphs," *arXiv preprint arXiv:1607.04842*, 2016.
[16] S. Li, M. A. Maddah-Ali, and A. S. Avestimehr, "Coded MapReduce," in *2015 53rd Annual Allerton Conference on Communication, Control, and Computing (Allerton)*, 2015, pp. 964–971.
[17] K. Lee, M. Lam, R. Pedarsani, D. Papailiopoulos, and K. Ramchandran, "Speeding up distributed machine learning using codes," *arXiv preprint arXiv:1512.02673*, 2015.
[18] M. Attia and R. Tandon, "Information theoretic limits of data shuffling for distributed learning," *arXiv preprint arXiv:1609.05181*, 2016.
[19] L. Song, C. Fragouli, and T. Zhao, "A pliable index coding approach to data shuffling," in *2017 IEEE International Symposium on Information Theory (ISIT)*, 2017, pp. 2558–2562.
[20] M. A. Maddah-Ali and U. Niesen, "Fundamental limits of caching," *IEEE Transactions on Information Theory*, vol. 60, no. 5, 2014.

# APPENDIX A
## PROOF OF THEOREM 2

In the low probability regime $p \leq p_{th}(m,n,K)$, we saw from Lemma 2 that some functions cannot be calculated directly after the random data placement. Therefore, we need to make transmissions in order to complete the computation tasks. For uncoded transmissions, we have two broadcast transmission scenarios: one is to make uncoded broadcast transmissions of the raw data messages $b_j$, $j \in [m]$, and the other is to make uncoded broadcast transmissions of the intermediate data messages $v_{k1}^k$, $v_{k2}^k$, $k \in [K]$[3]. Let us denote by $T_{un}$ and $T'_{un}$ the minimum number of transmissions needed for the above two scenarios so that $K$ nodes have all the data needed to compute the $K$ functions without further communication. It is not hard to see that $T'_{un} \geq T_{un}$ since a single raw data messages can be used to generate several intermediate data messages. We are going to use $T_{un}$ as a benchmark to compare the savings of broadcast transmissions using coding.

$T_{un}$ can be calculated using a bipartite graph matching problem: we list the $n$ nodes and the $K$ functions on both sides and connect a node to the functions it can compute. We then interpret $T_{un}$ as the minimum number of uncoded broadcast transmissions nodes need to make such that there is a matching that covers all $K$ functions. Clearly, if after the random data placement, we are only allowed to exchange intermediate function values, $T_{un}$ would form a lower bound on the number of uncoded transmissions needed $T'_{un}$, since broadcasting raw data could be useful to several nodes, while intermediate values are only useful to a single node. We are now going to show that, even if we are allowed to make coded broadcast transmissions, we would not achieve significant transmission savings (in order of magnitude).

We first note that, we have at least $T_{un}$ nodes missing at least $T_{un}$ different data. The number of coded transmissions we make need to satisfy at least these nodes. Using a similar approach as in index coding [3], we can define a matrix $G \in \mathbb{F}_q^{T_{un} \times T_{un}}$ that "fits" our data shuffling problem as follows:
- the columns are indexed by $T_{un}$ missing messages - we assume without loss of generality these to be the first $T_{un}$, $j = 1 \ldots T_{un}$ out of the $m$ messages;
- the rows are indexed by some $T_{un}$ computing nodes, $i_1, i_2, \ldots, i_{T_{un}}$, with node $i_j$ missing data $j$ for some function calculation, i.e., $j \notin S_{i_j}$;
- the $(i,j)$ entry of $G$, $g_{i,j} = 1$ for $i = j$, $g_{i,j} = 0$ for $j \in [m] \setminus S_i$, and $g_{i,j}$ can be arbitrary element in $\mathbb{F}_q$ for $j \in S_i$;
- node $i_j$ can use row $i_j$ to decode the message $j$ it needs.

Similarly to [3], let $\mathcal{G}$ denote the set of all matrices that fit our problem, the minimum number of coded broadcast transmissions equals the minimum rank of any matrix in $\mathcal{G}$, denoted as $minrank(\mathcal{G})$.

---
[3]The first scenario to transmit the raw data messages can be seen as a special case of transmitting the intermediate data messages.

---

To derive a lower bound on $minrank(\mathcal{G})$, we use a similar technique as in [15]. Let the rank of $G$ be $k$. For a principal submatrix $M$ of $G$, we denote the weight (number of nonzero elements) of $M$ by $w'$, the minimum weight of a column basis of $M$ by $w'_C$, and the minimum weight of a row basis of $M$ by $w'_R$. Let the rank of $M$ be $k'$. We define a coding structure $M$ to be a $n' \times n'$ principal submatrix of $G$ with the following properties [15]:

- the weight and the rank of $M$ satisfy $4k'w' \geq n'^2$, $k'/n' \leq k/T_{un}$, $w'_C \leq 2w'k'/n'$, and $w'_R \leq 2w'k'/n'$;

In [15], it is shown that every matrix $G$ contains such a coding structure $M$ and the number of such principal submatrices is at most $(n'q)^{12w'k/T_{un}}$.

The probability that a matrix $G$ containing $M$ is a fitting matrix is $p^{w'-n'}$, since we need that side information covers all off-diagonal non-zero elements of $M$. The number of ways we choose an ordered set of $n'$ computing nodes for submatrix $M$ is $n(n-1)(n-2)\ldots(n-n'+1) \leq n^{n'}$. The number of ways of choosing $M$ among all possible submatrices of $G$ is $\binom{T_{un}}{n'}$. By setting $k = \Theta(T_{un})$, we can calculate

$$\Pr\{minrank(\mathcal{G}) \leq k\}$$
$$\leq \sum_{k',n',w'} n^{n'} \binom{T_{un}}{n'}(n'q)^{12w'k/T_{un}} p^{w'-n'}$$
$$\leq \sum_{k',n',w'} 2^{n'\log(n)+n'\log(T_{un})+12w'k/T_{un}\log(n'q)+(w'-n')\log(p)}$$
$$\leq n^4 2^{(2.25n'+12w'k/T_{un})\log(nq)-w'/4\log(n)}$$
$$\leq n^4 2^{14.25w'k/T_{un}\log(nq)-w'/4\log(n)} = 2^{-\Omega(w'\log(n))},$$

where the first three inequalities follow from the properties of $M$ and $p \leq \sqrt{\frac{\log(K)}{n}}$, and the last equality follows from $k = \Theta(T_{un})$ and $w' \geq C$ for some large enough fixed constant $C$ according to the properties of $M$.